\documentclass{optica-article}

\journal{opticajournal} % for journals or Optica Open

\articletype{Research Article}

\usepackage{lineno}
\begin{document}

\title{Light trapping using dimer of spherical nanoparticles based on titanium nitride for plasmonic solar cells}
\author{Nowshin Akhtary,\authormark{1} Ahmed Zubair\authormark{1,*}}

\address{\authormark{1} Department of Electrical and Electronic Engineering, Bangladesh University of Engineering and Technology, Dhaka 1205, Bangladesh \\
}

\email{\authormark{*}ahmedzubair@eee.buet.ac.bd}  

% use {asbstract*} to suppress the copyright line. Copyright information will be added in production

\begin{abstract*} 
Light-trapping mechanisms with plasmonics are an excellent way to increase the efficiency of photovoltaics. Plasmonic dimer-shaped nanoparticles are effective in light absorption and scatterings, and there is hardly any research on dimer TiN nanoparticle-based PV. This paper demonstrated that titanium nitride could be a suitable substitute for other plasmonic materials in the visible and near-infrared spectrum. We designed a TiN-based spherical dimer plasmonic nanoparticle for photovoltaic applications. We conducted comparison analyses with the metals  Ag, Au, and Al to ascertain the performance of TiN as a plasmonic material. Silicon had an average absorption power of $\sim$19$\%$, and after incorporating TiN nanoparticles, the average absorbed power increased significantly to $\sim$75$\%$ over the whole spectral range. TiN dimer nanoparticle had the highest absorption cross-section, $Q_{ab}$ value $\sim$6.2 W/m$^2$ greater than Ag, Au, and Al had a fraction of light scattered into the substrate value greater than Au, Al and comparable to Ag. TiN dimer exhibited better absorption enhancement, $g$ for the whole spectral range than Ag, Au, and Al dimers for a radius of 15 nm with a peak value greater than 1. The maximum optical absorption efficiency of the plasmonic TiN nanostructures was $\sim$ 35.46$\%$.
\end{abstract*}

%%%%%%%%%%%%%%%%%%%%%%%%%%  body  %%%%%%%%%%%%%%%%%%%%%%%%%%
\section{Introduction}
Coal, natural gas, oil, biomass, and nuclear energy are non-renewable energy sources that are becoming scarcer and more depleted every day. Therefore, the abundance, affordability, and low environmental impact of renewable energy sources are all very advantageous. The efficiency of photovoltaic (PV) cells powered by renewable energy sources has been a great focus of research\cite {green2003impurity, enrichi2018plasmonic}. Light absorption and the frequency of electron-hole pair formation have a significant role in how well PV cells perform. The most common PV cells in CMOS technology are based on a silicon absorber layer with the limitation of high production costs and thinner absorption volume. There are many ways of increasing efficiency, such as surface texturing\cite{kim2020surface, yan2012correlation, ren2015topology}, metal nanograting \cite{jung2020polarization,dhawan2011narrow}, tandem structure \cite{bailie2015high}, optical absorption enhancement by increasing the effective optical path length or trapping light in the cell by introducing light scatters in the solar cell \cite{tamang2016enhanced,zhou2015plasmon,wang2016highly,atwater2011plasmonics,catchpole2008design, yu2018giant, Akhtary2023}. A suitable choice of materials for active layers can ensure better photon absorption, generating electron-hole pairs. However, only efficient absorption cannot generate efficient electron-hole pairs and, consequently, photo-voltage. The recombination process creates a loss of charge carriers; therefore, an optically thick semiconductor is unsuitable for better charge carrier separation. Additionally, more materials are needed for thicker semiconductors, which is cost-ineffective and wasteful. Thus, a thinner semiconductor layer is preferred.

Introducing metallic nanoparticles (NPs) has created an alternative approach to improving absorption efficiency. Surface plasmon resonance (SPR) can significantly enhance EM waves by placing plasmonic structures in the active layer. This phenomenon ensures enhanced light absorption providing strong scattering between the intense plasmon field and the active layer\,\cite{zhou2015plasmon, xue2011charge}. Localized surface plasmon resonances (LSPRs) generate light scattering by NPs. The LSPR occurs when the frequency of the optical photon coincides with the natural frequency of the collective vibration of conduction electrons in NPs, leading to strong near-field electromagnetic enhancement, acute spectral absorption, and scattering peaks \cite{mayer2011localized}. The enhancement of optical absorption of NPs is the foremost attribute of LSPR \cite{ihara1997enhancement}. Light absorption and photo-current have been improved by using the LSPR phenomena \,\cite{baba2011increased,standridge2009distance}. Much recent research has centered on the trade-off between optimal thickness and maximum field enhancement. The most significant improvement variables ordinarily happen when the junction between the absorber and NPs is illuminated by polarized light. Complex structures can achieve a sensitivity that leads to near-infrared (NIR) sensing and plasmon hybridization. The NPs structure can get over this restriction since it extends the inside field to the outside environment, which results in a considerable boost in detection sensitivity. 

Plasmonic materials can support electrons or plasmons across a broad spectrum from infrared to ultraviolet solar radiation. Until recently, researchers were confined to noble metals like Ag and Au as plasmonic material. Ag and Au are frequently used plasmonic metals and optical metamaterials because of their strong DC conductivity or low resistive losses. While an electron in a metal's valence band absorbs a photon to jump to the Fermi surface or while an electron close to the Fermi surface absorbs a photon to fill the ensuing unoccupied conduction band, there is confinement for plasmonic metals, causing an excessive loss in conventional plasmonic materials. Ordinary metals have several drawbacks, including the size of the genuine portion of the permittivity, the ineffectiveness of tuning or balancing the optical properties of metals, and their high cost.

Due to the high optical loss of metals, alternative metals with the least ohmic loss may be preferred for plasmonic devices. To reduce the interband transition loss, many reports frequently utilized alternative plasmonic NP\,\cite{cai2010optical}. Conventional plasmonic materials have many shortcomings, leading researchers to seek better alternatives. The alternative plasmonics has the real permittivity of the same order. Hence, geometric fractions lights can be promptly tuned to coordinate the plan prerequisites. Conventional plasmonic metals confront debasement when exposed to air/oxygen or moisture, causing further problems in device fabrication and integration. These criteria directly affect optical properties and increase optical loss, resulting in more significant values of the dielectric function's imaginary part and rendering it incompatible with conventional silicon fabrication methods.  Metal nitrides are a better alternative to overcome the shortcomings. Among them, titanium nitride (TiN) is a non-stoichiometric, interstitial compound with a high concentration of free carriers. It is refractory and steady, and its optical properties can be tuned by changing its geometric structure\,\cite{Shariful2023}. Moreover, it is consistent with silicon CMOS technology \cite{mandal2016progress,gangadharan2017recent} and offers manufacturing and integration advantages that can help overcome the challenges. There are several reports on monomer spherical and hemispherical TiN NPs \cite{khalifa2014plasmonic, naik2013titanium}. However, no dimer spherical TiN NP-based plasmonic solar cells have been reported.

This paper employed the finite-difference time-domain (FDTD) method to systematically investigate the scattering cross-section and absorption enhancement by spherical dimer TiN NPs for photovoltaic application.In order to ascertain the total scattering cross-section, the percentage of light scattered into the substrate, the absorption cross-section, and the spatial mapping of the electric field in this plasmonic nanosystem, we first built and optimized the dimer of the spherical NPs. We further investigated the effect of polarization sensitivity on the source. We gained insight into how the shape of the NPs enhanced the functionality of solar cells when the NPs were embedded into them. We investigated plasmonic core-shell configuration and analyzed the effect of dielectric coatings on NPs. Our work provided insights into using TiN in photovoltaic cells.

\section{Methodology}

\begin{figure}[h]
\hspace*{-1.2cm}
\includegraphics[trim={0cm .05cm 0cm .1cm},clip,width=1.05\textwidth]{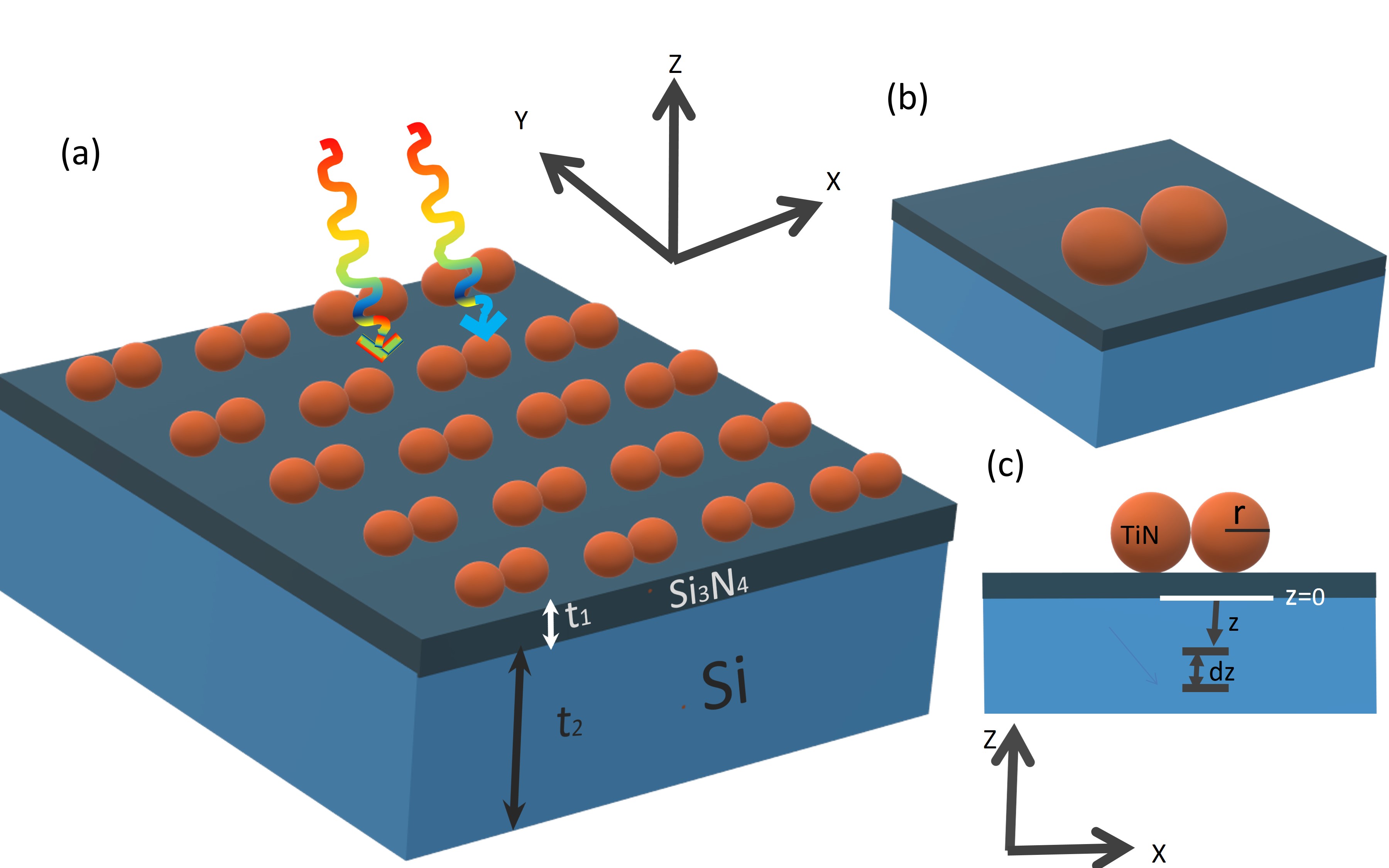}
\caption {(a) Illustration of  spherical dimer TiN nanoparticle on a 30 nm thin Si$_3$N$_4$ underlayer on Si substrate acting as the absorber layer of the photovoltaic cell. (b) Perspective view and (c) cross-sectional view of a single dimer of spherical nanoparticles.}
\label{fig}
\end{figure}
 
\begin{figure}[h]
\hspace*{-2.cm}
\includegraphics[trim={0cm .5cm 0cm .4cm},clip,width=1.25\textwidth]{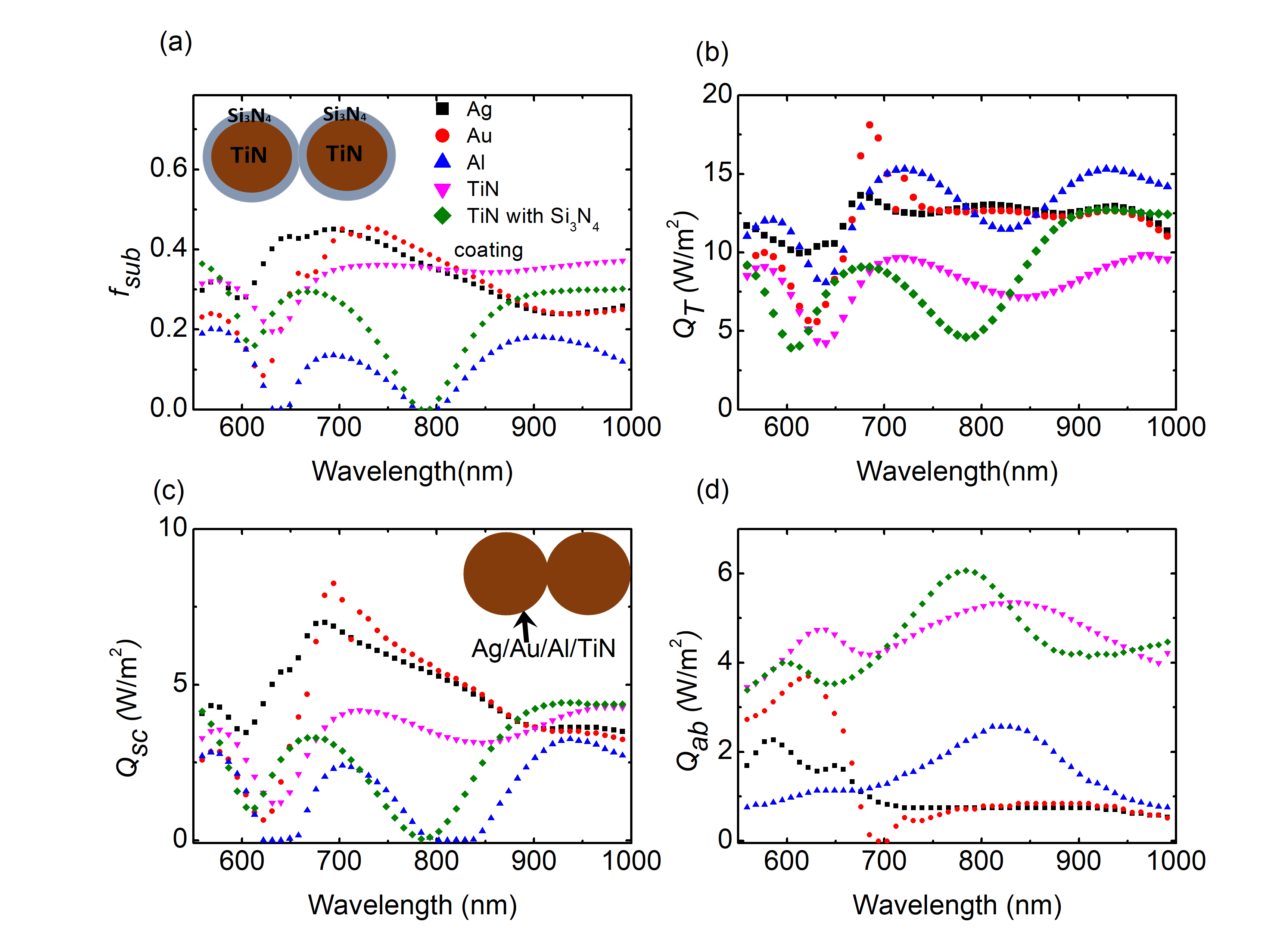}
\caption {(a) $f_{sub}$, (b) $Q_{T}$, (c) $Q_{sc}$, and (d) $Q_{ab}$ as a function of wavelength for 100 nm radius dimer NPs  consisting of Ag, Au, Al, TiN, and TiN with Si$_3$N$_4$coating on a 30 nm thick Si$_3$N$_4$ underlayer on Si substrate.}
\label{dimer}
\end{figure}

\begin{figure}[h]
\hspace*{-1.8cm} 
\includegraphics[trim={0cm 0cm 0cm 0cm},clip,width=1.25\textwidth]{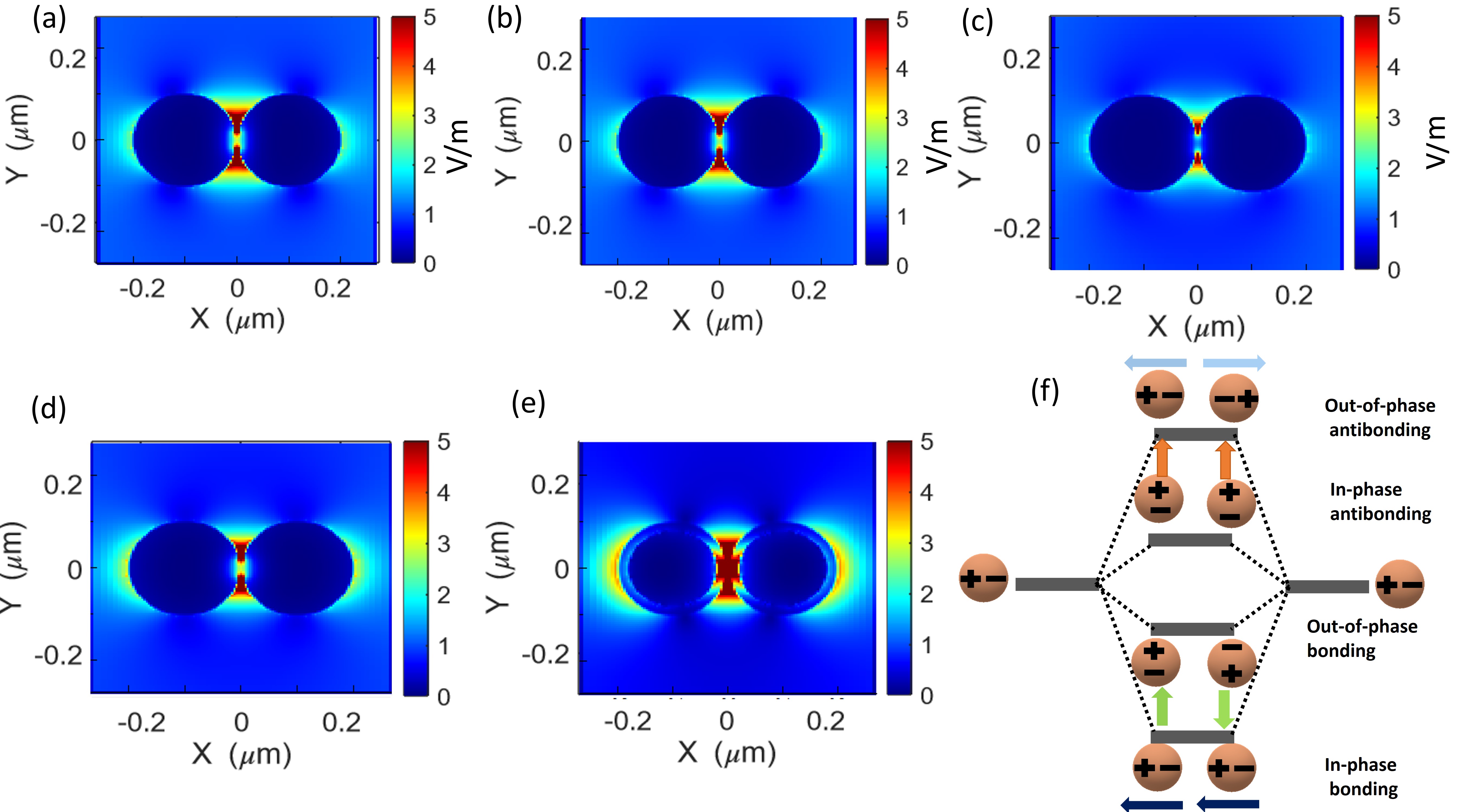}
\caption {Color map showing the xy distribution of |E| for (a) Ag, (b) Au, (c) Al (d) TiN (e) (d) TiN with Si$_3$N$_4$ coating dimer NP placed on top of a 30 thin Si$_3$N$_4$ on a Si substrate. (f) Schematic of the LSPR plasmon modes for NP dimers. The coupling of the dipoles in the two spheres of the dimer, created by the plasmon, can occur in the direction of the dimer axis or perpendicular to it.}
\label{metal_z}
\end{figure}

\begin{figure}[h]
\hspace*{-1.8cm}
\includegraphics[trim={0cm .3cm 0cm .7cm},clip,width=1.25\textwidth]{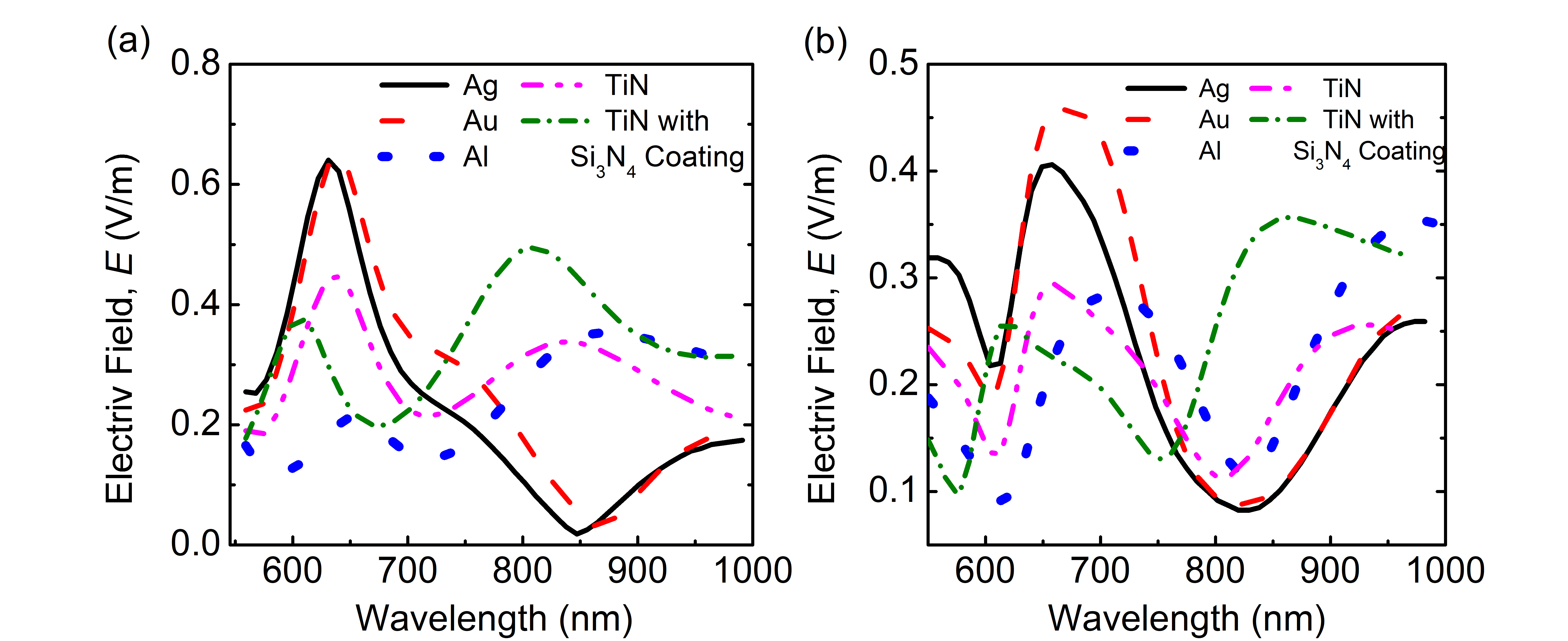}
\caption {The E-field spectra at (a) yz, and (b) xy plane for spherical dimer NPs comprised of different materials placed on top of a 30 thin Si$_3$N$_4$ on a Si substrate.}
\label{Efield}
\end{figure}

\begin{figure}[h]
\hspace*{-1.8cm} 
\includegraphics[trim={0cm 0.5cm 0cm 0cm},clip,width=1.25\textwidth]{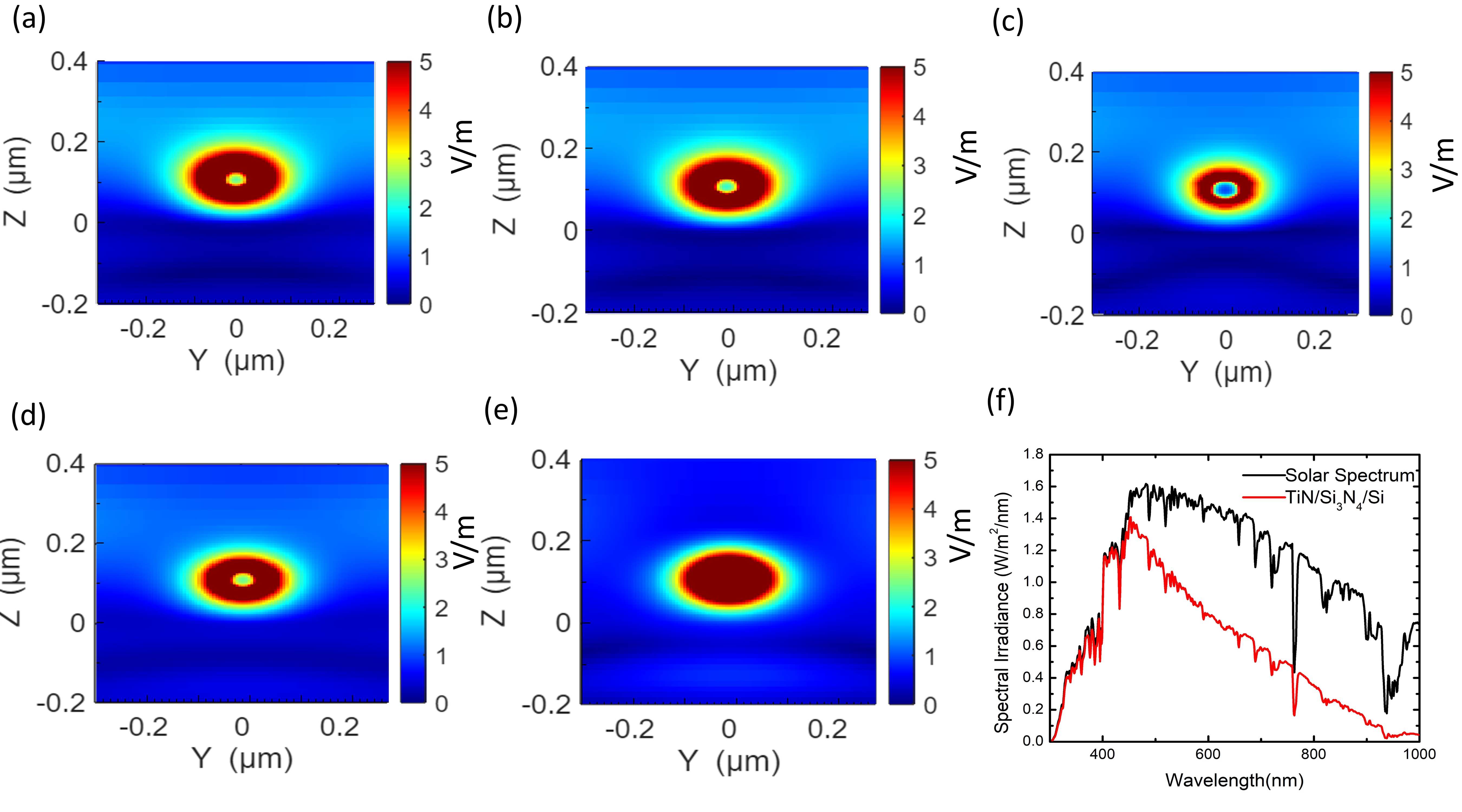}
\caption {Color map showing the yz distribution of |E| for (a) Ag, (b) Au, (c) Al (d) TiN (e)  TiN with Si$_3$N$_4$ coating dimer NP placed on top of a 30 thin Si$_3$N$_4$ on a Si substrate. (f) Absorption spectra of TiN dimer NP incorporated Si absorber layer and Solar irradiance spectra.}
\label{metal_x}
\end{figure}

\begin{figure}[h]
\hspace*{-1.8cm}
\includegraphics[trim={0cm 1.7cm 0cm .7cm},clip,width=1.25\textwidth]{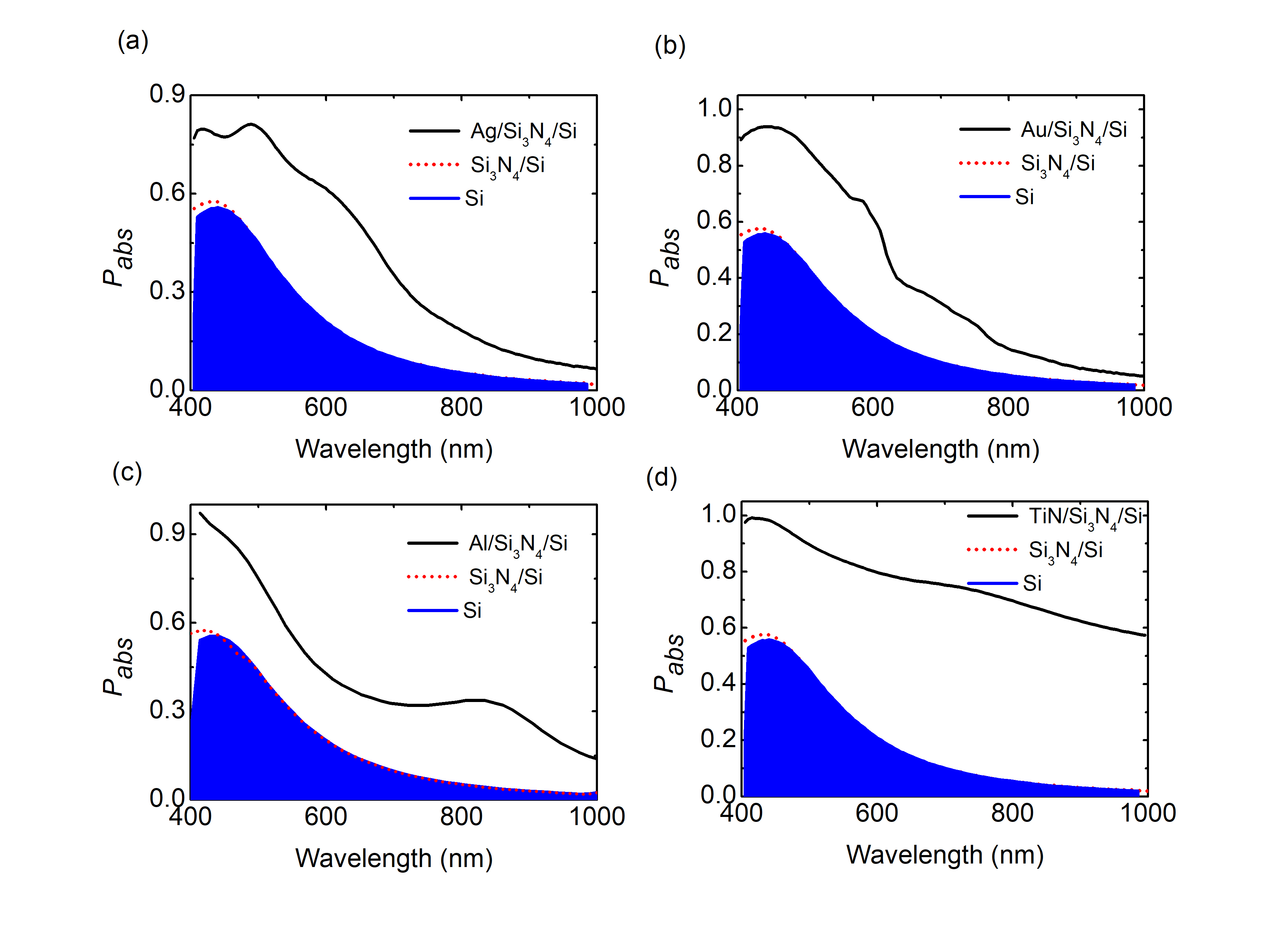}
\caption {Percentage of power absorbed for (a) Ag, (b) Au, (c) Al, (d) TiN NP incorporated Si absorber layer for solar cells.}
\label{power}
\end{figure}

\begin{figure}[h]
\hspace*{-2.cm}
\includegraphics[trim={0cm .5cm 0cm .4cm},clip,width=1.25\textwidth]{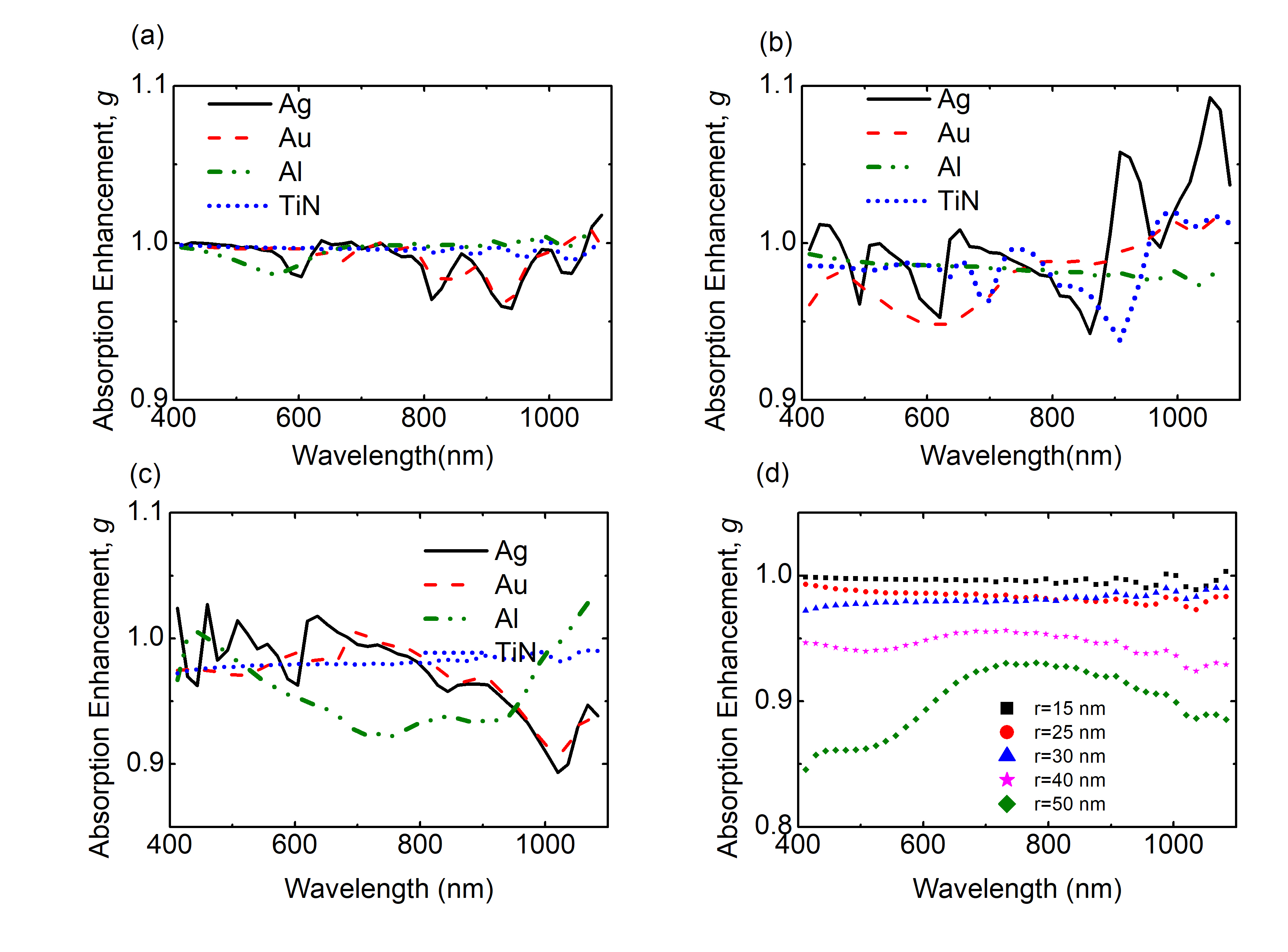}
\caption {Absorption enhancement, $g$ for the dimer of plasmonic spherical NPs for r = (a) 15 nm, (b) 25 nm, and (c) 30 nm. (d) The $g$ for TiN NPs with different radii.}
\label{enhancement}
\end{figure}

\begin{figure}[h]
\hspace*{-2.2cm} 
\includegraphics[trim={0cm 1.7cm 0cm .4cm},clip,width=1.25\textwidth]{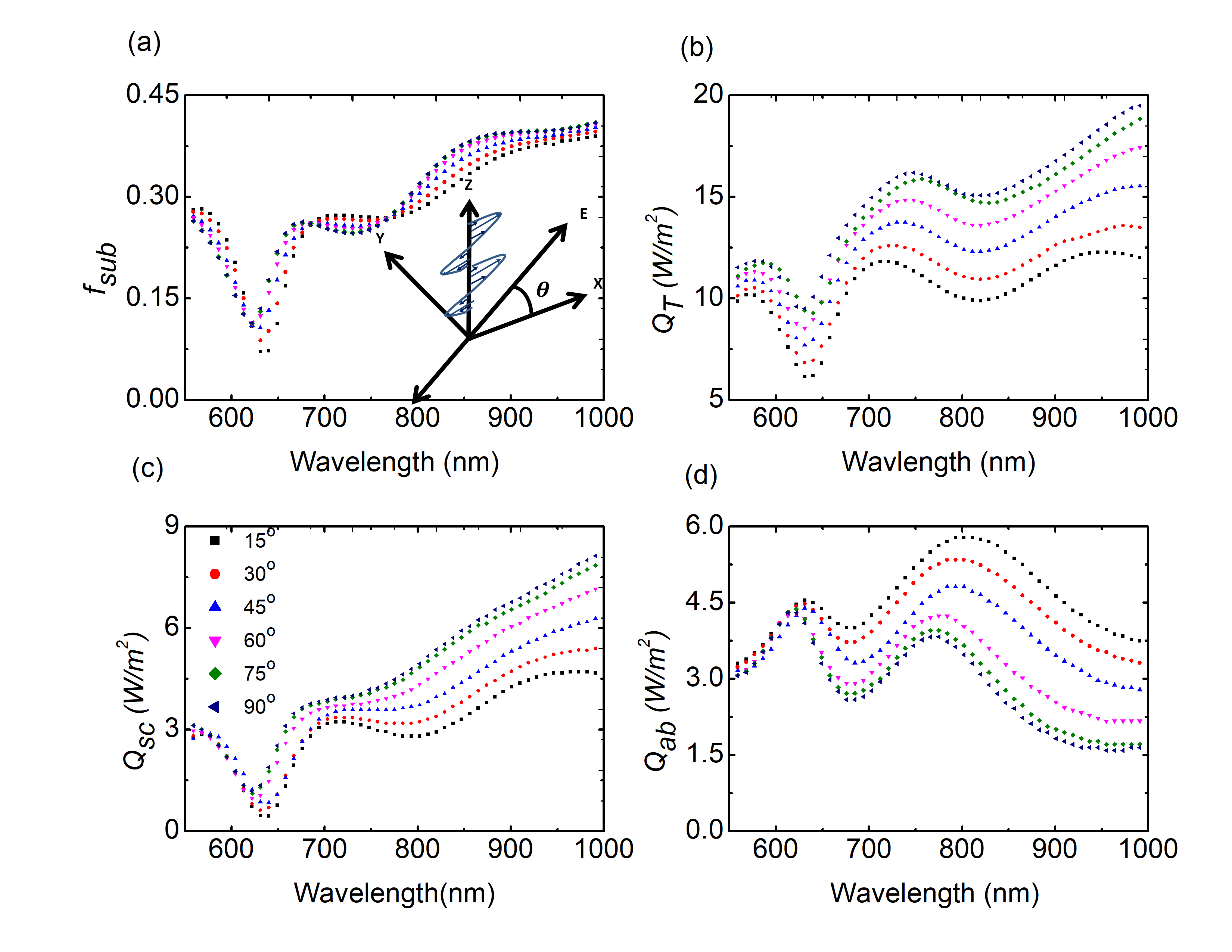}
\caption {(a) $f_{sub}$, (b) $Q_{T}$, (c) $Q_{sc}$, and (d) $Q_{ab}$ as a function of wavelength for TiN dimer shaped NP with varying the polarization angle of the source from $\theta$=15$^\circ$ to 90$^\circ$.}
\label{angle}
\end{figure}

\begin{figure}[h]
\hspace*{-1.8cm} 
\includegraphics[trim={0cm 0.5cm 0cm 0cm},clip,width=1.25\textwidth]{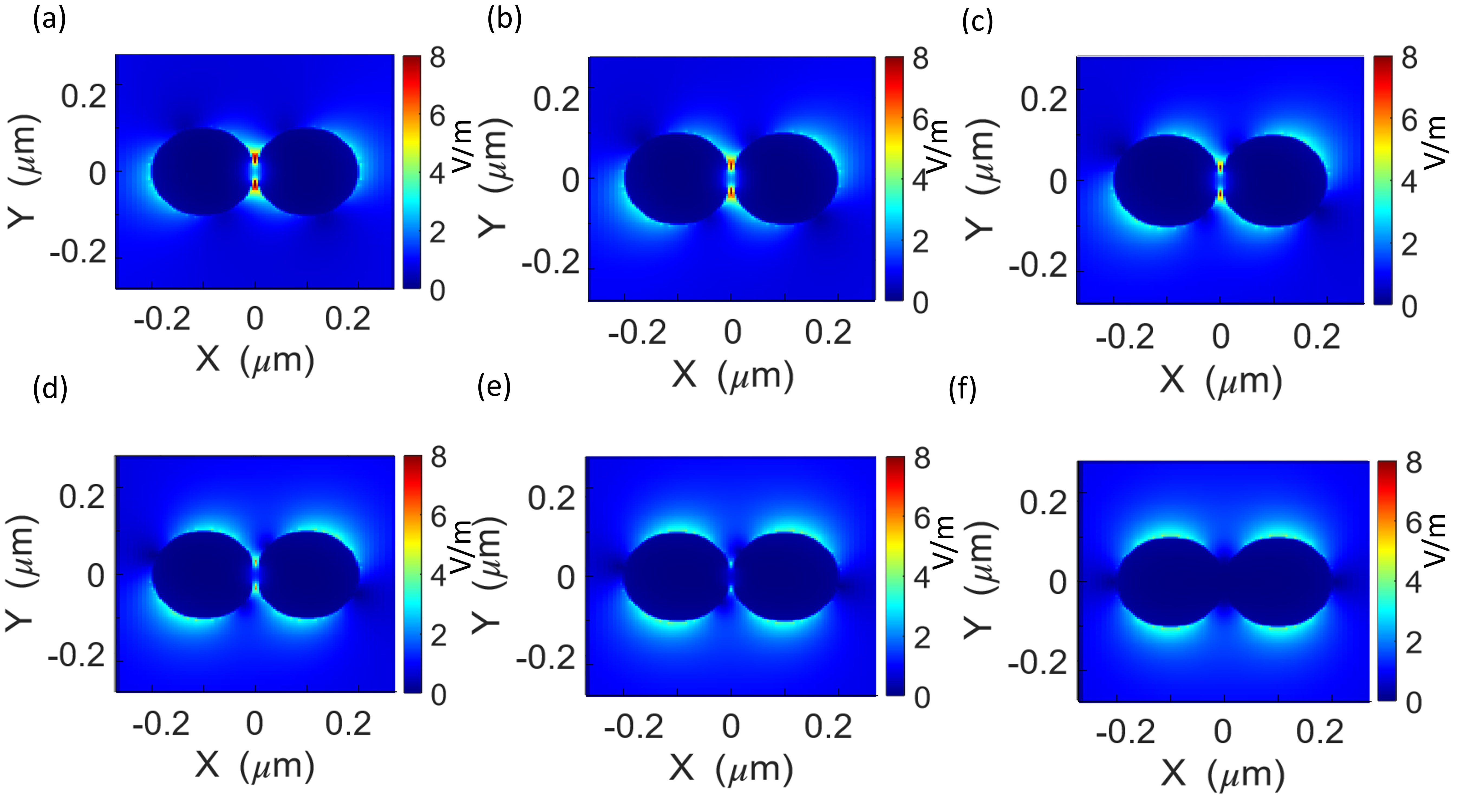}
\caption {Color map showing the distribution of E-field in xy plane at Z=0 for TiN dimer NP placed on top of a 30 nm thin Si$_3$N$_4$ on a Si Substrate with varying the polarization angle of the source from $\theta$ = 15$^\circ$ to 90$^\circ$.}
\label{colormap_po}
\end{figure}

\begin{figure}[h]
\hspace*{-1.8cm} 
\includegraphics[trim={0cm 0.5cm 0cm 0cm},clip,width=1.25\textwidth]{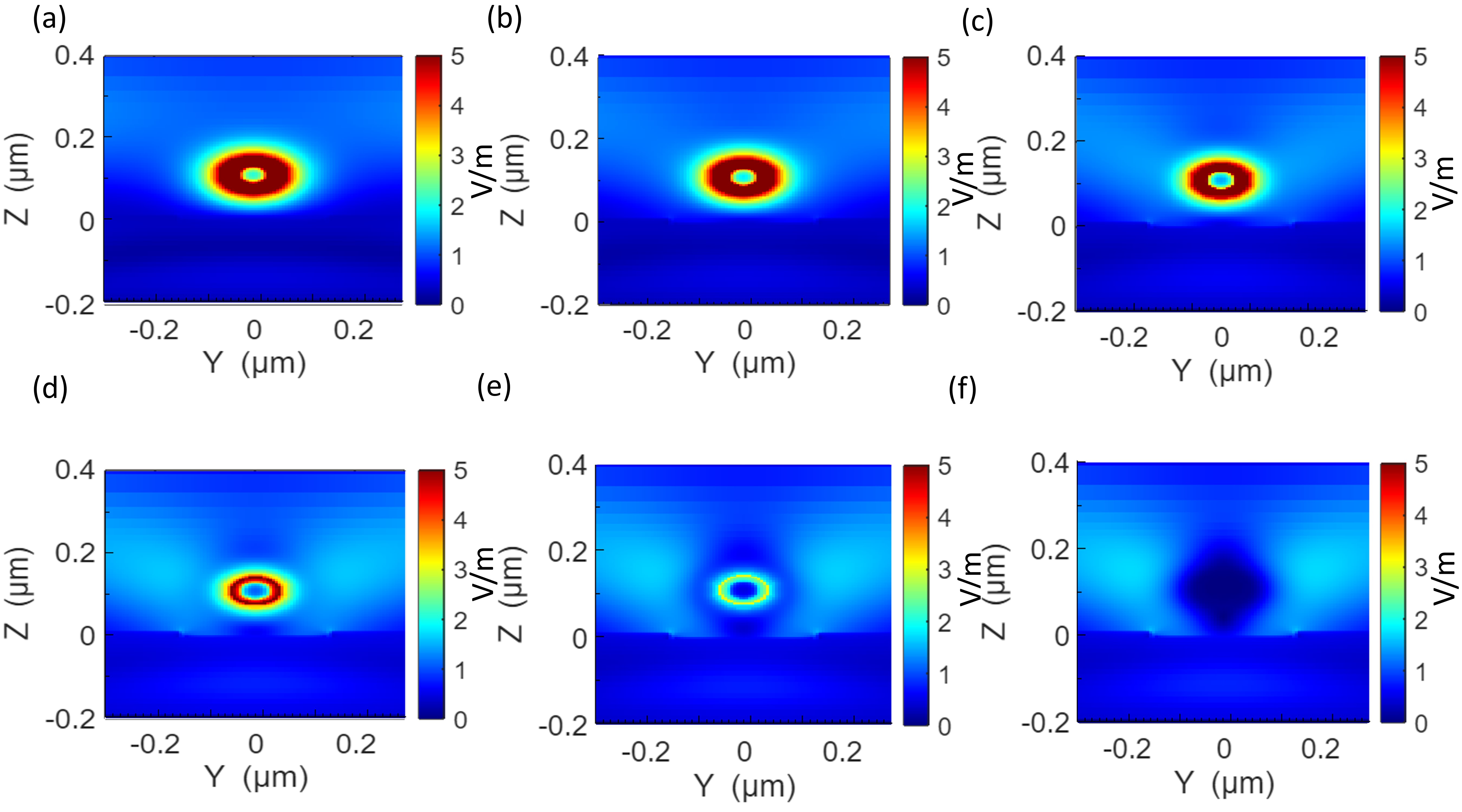}
\caption {Color map showing the distribution of |E| in the yz plane at x=0 for TiN dimer NP placed on top of a 30 nm thin Si$_3$N$_4$ on a Si Substrate with varying the polarization angle of the source from $\theta$ = 15$^\circ$ to 90$^\circ$.}
\label{colormap_po1}
\end{figure}

\subsection{Structural Design}
We developed an alternative plasmonic material, TiN-based spherical dimer NP on a semi-infinite crystalline silicon substrate as can be seen in Fig.\,\ref{fig} (see Fig.S1 of \textcolor{blue}{Supplement 1}). In the visible and longer wavelengths, TiN displays localized surface plasmon phenomena and metallic characteristics. \cite{naik2012titanium,naik2014alternative}. The plasmonic particles were separated from the semi-infinite silicon absorption layer by a thin Si$_3$N$_4$ layer as surface passivation. We compared cross-sections of NPs based on conventional noble plasmonic metal with TiN alternative plasmonic NPs. The size of the particles was varied, and their properties were analyzed. t$_1$ and t$_2$, respectively, presented the thin film and substrate thickness. The source's polarization angle represented by $\theta$, r represented the radius of the sphere, and d represented the distance between the nanospheres of a dimer.

\subsection{Simulation Methods}
We applied the FDTD method, where Maxwell's equations were solved numerically, to study the mentioned nanosystems. The simulation dimensions of the FDTD were 1.2 $\mu$m $\times$ 1.2 $\mu$m $\times$ 1.25 $\mu$m. A mesh size of 0.4 nm  was applied around the NPs. The source was adjusted for polarization perpendicular to the surface normal of the particles from the air side. The particles were incident to the total-field scattered-field (TFSF) plane wave along the negative z-axis. The incident source was a uniform wave with a prominent wavelength range of 550–1100 nm, which comprised the solar spectrum's highest feasible irradiance (AM 1.5). A plane wave with a TFSF was utilized to separate the incident field from the scattered field to examine the optical characteristics of NPs. The scattering characteristics were investigated using an external monitor. The spatial electric field mapping was performed by adjusting a frequency-domain power monitor. We used light scatterers to increase the light trapping efficiency, improving the absorber layer's absorption. We estimated the scattering and absorption cross-sections, and optimal values were obtained for better PV application by adjusting various factors. The electric and magnetic fields around the particle were calculated by converting the time domain into the frequency domain using a Fourier transform. The radial Poynting vector, $S(\omega)$, was calculated from the electric field, $E(\omega$), and magnetic field $H(\omega)$ as a function of angular frequency, $\omega$. In the scattered field region, the total of power ${P_{s}}$ was determined along the +x, +y, +z, --x, --y, and --z directions. The ratio of the power in the scattered field region inside the substrate of the absorber layer to the power in the scattered field region in the air and the absorber layer is known as the percentage of light scattered into the substrate, $f_{sub}$. The total scattering cross-section, $Q_{T}(\omega)$ is defined as the sum of the power per unit area scattered in all directions divided by the power per unit area of the incident beam. 
\begin{equation}\label{C}
Q_{T}(\omega)= \frac{P_{s}(\omega)}{I(\omega)}.
\end{equation}
Here, $I(\omega)$ is incident power intensity as a function of $\omega$. Absorption cross-section is a measure of the probability of an absorption process. The total absorbed power divided by the power per unit area of the incident light was defined as absorption cross-section, $Q_{ab}$. It can be calculated from
\begin{equation}
dN/dz=-NnQ_{ab}
\end{equation}
Here, dN/dz is the number of photons absorbed between the points z and z+dz, N is the number of photons penetrating to depth z, and n is the number of absorbing molecules per unit volume. The monitors outside the TFSF source determined the scattering cross-section.\,\cite {catchpole2008design,  ye2017plasmonic}. 

We methodically considered the impact of NPs' $Q_{T}$, the light scattered into the substrate $Q_{sc}$, $f_{sub}$, and $Q_{ab}$ by investigating their structural characteristics. We compared the effects of alternative plasmonic NPs to plasmonic metals and comprehensively considered the capacity to enhance absorption within an absorber layer by adding NPs. Moreover, to demonstrate the effectiveness of the dimer NPs in PV cells, we calculated the proposed structure's absorption enhancement and light absorption efficiency. 

\section{Results and discussion}

\subsection{Effect of different material-based plasmonic spherical dimer  nanoparticle}
We simulated  spherical dimer NPs for different materials and tracked their scattering and absorbance behavior. Here, t$_1$, t$_2$, and r were regarded as 30 nm, 250 nm, and 100 nm, respectively. We evaluated the performance of NPs made of different materials, including Au, Ag, Al, and TiN. Moreover, we explored a core-shell configuration consisting of TiN NP with Si$_3$N$_4$ coating to maximize the performance. The foremost critical factor representing the path length enhancement of a scattering light-trapping structure is $f_{sub}$\,\cite{bohren2008absorption}. As can be seen in Fig.\,\ref{dimer} (a), the overall $f_{sub}$ exhibited Ag> Au> TiN> Tin with Si$_3$N$_4$ coating> Al in this order. For 800 nm to 1100 nm wavelength, the$f_{sub}$ of TiN NP was greater than those for Ag, Au, and Al-based NPs. When we varied the dimer materials, the peak value of $Q_{T}$ was 19 W/m$^2$ for Au NP, and the values for Ag and Al were comparable to Au. TiN NP had comparable values of $Q_{T}$ and $Q_{sc}$ from 650 to 1000 nm wavelength. After adding Si$_3$N$_4$ coating on TiN NP the $Q_{T}$ and $Q_{sc}$ performed better for 850 to 1000 nm as can be seen Fig.\,\ref{dimer}(b)-(d). For scattering applications, Si$_3$N$_4$ coating can be used for TiN NP. $Q_{ab}$ increased for TiN and TiN with Si$_3$N$_4$ coating NPs. The $Q_{ab}$ was negligible for Ag, Au, and Al NPs.

When a plane wave collides with an object or scatterer, its energy is diverted in all directions. It is crucial to analyze the optical properties of NPs, including scattering cross-section and electric field distribution. By changing the spherical dimer NPs' material, shown in Fig.\,\ref{metal_z} electric field maps in the xy plane were detected. LSPR modes can be produced by these structures. The plasmonic NP dimers are the equivalent of two atoms sharing electrons by bonding molecular orbitals. A dimer's excited dipoles on its two spheres may couple in the direction of the dimer axis, which is analogous to $\sigma$-type orbital for atoms or perpendicular to it, which is analogous to $\pi$-type orbital for atoms. Four additional plasmon modes emerge for dimers, which are homonuclear diatomic molecules equal to molecular hydrogen(H$_2$), nitrogen (N$_2$), oxygen (O$_2$), or a halogen (X$_2$)\cite{pascale2019full}. As shown in Fig.\,\ref{metal_z}(f), when the charge of the dimers oscillates in the same direction, the charge accumulates, and electric field enhancement is observed. This phenomenon occurs both in bonding mode and anti-bonding mode, and they are in-phase antibonding with the highest energy and in-phase bonding with dipolar plasmon mode with the lowest energy (highest wavelength) (see Fig.\,\ref{metal_z}(f)). Additionally, when the charge oscillates in different directions, there is no field enhancement, which is out-of-phase bonding and antibonding modes\,\cite{schumacher2019precision}. As can be seen in Figs.\,\ref{metal_z}(d)-(e), the scattering spectra TiN and TiN with a dielectric Si$_3$N$4$ coating exhibited an unprecedented homogeneity for the two spheres. The in-phase bonding plasmon mode was observed for x-polarized light %at highly uniform positions along the x-axis with a very narrow standard deviation. %For Ag, Au, and Al NP had zero effective induced dipole moment along the corners of the y-axis causing dark mode. 
The induced dipole moments resulted in two bright modes. Therefore, the accumulation of high free charge distribution around the surface and center of the dimer resulted in the enhancement of the electric field.

The quasistatic dipole approximation was used to compute the electric field ($E$) enhancement at yz and xy plane around the surface of Ag, Au, Al, and TiN dimers presented in Fig\,\ref{Efield}.  Due to a lower real permittivity than Ag and Au, the magnitude of the magnetic field enhancement in TiN nanospheres was slightly smaller than those of Ag and Au. E-field intensity on the yz plane at x = 0, which is the center of the dimer, can be seen in Fig.\,\ref{metal_x} for different material-based dimer nanosphere. For Ag and Au and TiN with Si$_3$N$4$ coating, charge distribution was high at the center of the dimers as presented in Figs.\,\ref{metal_x}(a), (b) and (e). For Al and TiN, charge distribution at the center was less as compared to Ag and Au, as can be seen in Figs.\,\ref{metal_x}(c) and (d). Here, the charge oscillates in the same direction, so the charge accumulated and filed enhancement occurred.  Resulting LSPR modes were utilized in numerous detection applications \cite{naik2012titanium}. The LSPR mode of TiN with Si$_3$N$4$ coating in the core-shell configuration was blued-shifted, and the peak value of the electric field increased compared to bare TiN dimer.

To determine the effectiveness of the NP in the photovoltaic cells, we calculated the absorbed power of each layer of nanostructure consisting of dimer spherical NPs on a 30 nm thin Si$_3$N$_4$ underlayer on Si substrate. NPs were comprised of Ag, Au, Al, and TiN. The divergence of the Poynting vector was used to compute the absorption per unit volume as given by,
\begin{equation}\label{abs1}
p_{abs}=-\frac{1}{2}\,real\,
(\vec{\nabla} \,.\,\Vec{S}).
\end{equation}
However, divergence calculations are frequently quite susceptible to numerical errors. Consequently, the simplest method for calculating absorbed power is,
 \begin{equation}\label{abs2}
   p_{abs}=-\frac{1}{2}\,real\,(i\omega\,\vec{E}. \Vec{D}^*).
\end{equation}
It can be modified as
 \begin{equation}\label{abs3}
 p_{abs}=-\frac{1}{2}\,\omega\,|E|^2\, imag(\epsilon).
 \end{equation}
Here, $D$ is the electric displacement field, and $\epsilon$ is the permittivity. Standard 1.5 ATM solar spectrum and absorption of TiN NP-based solar cell are presented in Fig.\ref{metal_x}(f). Solar light absorption is very efficient for wavelengths from 300 nm to 500 nm. For longer wavelengths than 500 nm, the absorption decreased gradually. Silicon had an average absorption power of $\sim$51$\%$ over the 400-500 nm range and $\sim$19$\%$ for the whole spectral range  as presented in Fig.\,\ref{power}. The Si$_3$N$4$ layer enhanced the absorption for the 400-500 nm spectral range. After the incorporation of TiN NPs, the absorbed power increased significantly to $\sim$75$\%$ over the whole spectral range as presented in Fig.\,\ref{power}(d). In comparison to Ag, Au, and Al NPs, TiN NP integration had a substantially higher absorption power.

\subsection{Absorption enhancement by TiN-based dimer NP}

TThe absorption enhancement defines the increased absorption by the addition of NPs in the solar absorber layer. The absorption enhancement spectrum, $g$ was presented  by,
\begin{equation}\label{abs}
    g(\lambda)=\frac{EQE_{np}(\lambda)}{EQE_{bs}(\lambda)}.
\end{equation}

Here, $EQE_{np}$ is the device's external quantum efficiency when plasmonic nanoparticles were incorporated on top of a substrate and $EQE_{bs}$ is the external quantum efficiency of a bare substrate. In this section t$_1$, and t$_2$ were taken to be 30 nm and 250 nm, respectively, in this section. We simulated the spectra of $g$ for Ag, Au, Al, and TiN plasmonic dimers on a silicon substrate for r = 15 nm, 25 nm, and 30 nm presented in Figs.\,\ref{enhancement}(a)-(c). TiN exhibited better absorption than Ag, Au, and Al dimers for r = 15 nm. For r = 25 nm the $g$ of TiN was $\sim$1 for the whole spectral range which was better than Au and Al. For r = 25 nm, the $g$ of TiN was between 0.9 to 1. For r = 30 nm the $g$ of TiN was better than Al and comparable to Ag, Au for 400 to 800 nm and the $g$ was greater than Ag, Au and Al for the range 800 nm to 1100 nm. For TiN, Al, Au, and Ag, the average enhancement, G, was discovered to equal 0.997, 0.991, 0.995, and 0.995, respectively.   

The light absorption efficiency (LAE) was calculated by,
%%%%%%%%%%%%%%%%%%%%%%
\begin{equation}
    \mbox{LAE} = \frac{\int_{400}^{1100} I(\lambda) A(\lambda)  d\lambda}{\int_{400}^{1100} I(\lambda) d\lambda}.
\end{equation}
Here, $I(\lambda)$ is the incident light intensity. And, absorbance, $A(\lambda)$ was calculated by, \begin{equation}
    A (\lambda) = 1 - R (\lambda) - T (\lambda).
\end{equation}
Here, $T(\lambda)$ and $R(\lambda)$ are the transmittance and reflectance of the structure. For the TiN plasmonic nanosphere on a kesterite substrate, we determined LAE. The values of LAE for r = 15 nm were found to be 35.46$\%$ and 33.78$\%$  for TiN and Ag respectively. 

We calculated $g$ for different radii of TiN plasmonic dimer on a silicon substrate as can be seen from Fig.\,\ref{enhancement}(d). The $g$ decreased with the increase of r for wavelength which agrees well with the previous study\,\cite{Akhtary2023}.This happened as a result of plasmonic NP's strong forward scattering and weak absorption of light at various radii. While backward scattering prevented absorption, forward scattering promoted it. Spherical dimer plasmonic NPs with larger radii often have larger cross-sectional areas of scattering, this increase of metal NP and higher arrange mode excitation can control light scattering which improves or diminish the light absorbing productivity into the substrate \cite{venugopal2017titanium}.

\subsection{Impact of light polarization}

As the polarization of light influences light scattering or absorption, we varied the polarization angle of the light source to compute the optical characteristics of the NPs\,\cite{Akhtary2023}. For the structure design with optimized scattering and absorption, we changed the source's polarization angle for dimer spherical TiN NPs and found the optimal polarization angle, $\theta$. Here, t$_1$ and r were regarded as being 30 nm and 100 nm, respectively. The $theta$ had been varied from 0$^\circ$ to 90$^\circ$. For the wavelength ranging from 550 to 760 nm, the value of $f_{sub}$ decreased with the increase of $\theta$ which is apparent from Fig.\,\ref{angle}(a). With a reduction in $\theta$, $f_{sub}$ remarkably dropped for the wavelength range of 760 nm to 1100 nm. As the $theta$ was raised, it was evident from Fig.\,\ref{angle}(b) that $Q_{T}$ increased significantly. The peak wavelength of $Q_{T}$ had a red-shifted spectrum. As can be seen from Fig.\,\ref{angle}(c), $Q_{sc}$ increased as the $\theta$ was increased and from Fig.\,\ref{angle}(d), $Q_{ab}$ increased as the angle decreased from 90$^\circ$ to 15$^\circ$.

Figs.\,\ref{colormap_po}(a)–(f) represented E-field intensity for dimer NP with the variation of $\theta$. The polarization angle defines the direction of the electric field and magnetic field. When the light was polarized in the x-direction (i.e., $\theta$=0), field enhancement was observed along the x-axis. As the angle changed, the induced dipole changed with $\theta$ and consequently, in-phase bonding and antibonding plasmon modes occurred accordingly\,\cite{schumacher2019precision, deng2018dark}. When $\theta$=15 $^\circ$ and the charge of the dimers oscillates in the same direction the charge accumulated, and electric field enhancement was observed at $\theta$=15 $^\circ$. This phenomenon occurred both in bonding mode and anti-bonding mode when they are in-phase plasmon mode. The direction of charge oscillation changed along the $\theta$ producing  an induced dipole moment\,\cite{Akhtary2023}. Fig\,\ref{colormap_po}(a)-(c) illustrated the strong electric field enhancement that was observed in the x-direction when charge oscillation was increasingly aligned to the x-direction. As the polarized angle rose from 45$^\circ$ to 90 $^\circ$  strong electric field enhancement was observed in the y-direction as charge oscillation alignment changed to y direction and the in-phase antibonding mode was observed as illustrated in Fig\ref{colormap_po}(d)-(f).

E-field intensity on the yz plane represented the center of the dimers, can be seen from Figs.\,\ref{colormap_po1}(a)-(f) with the variation of $\theta$. When $\theta$ increased from 15$^\circ$ to 45$^\circ$, the effective induced dipole moment decreased, which originated the in-phase antibonding mode, due to the E-field polarization alignment changed toward y direction presented in Figs.\,\ref{colormap_po1}(a)-(c). When $\theta$ increased from 60$^\circ$ to 90$^\circ$, the charge was distributed along the y-direction, which was the polarization angle of the E-field. Consequently, the out-of-phase antibonding mode was apparent when $\theta$ was near 90$^\circ$, as can be seen in Figs.\,\ref{colormap_po1}(d)-(f).

\subsection{Impact of the distance between the spheres of dimer}

\begin{figure}[h]
\hspace*{-2.cm}
\includegraphics[trim={0cm .5cm 0cm .4cm},clip,width=1.25\textwidth]{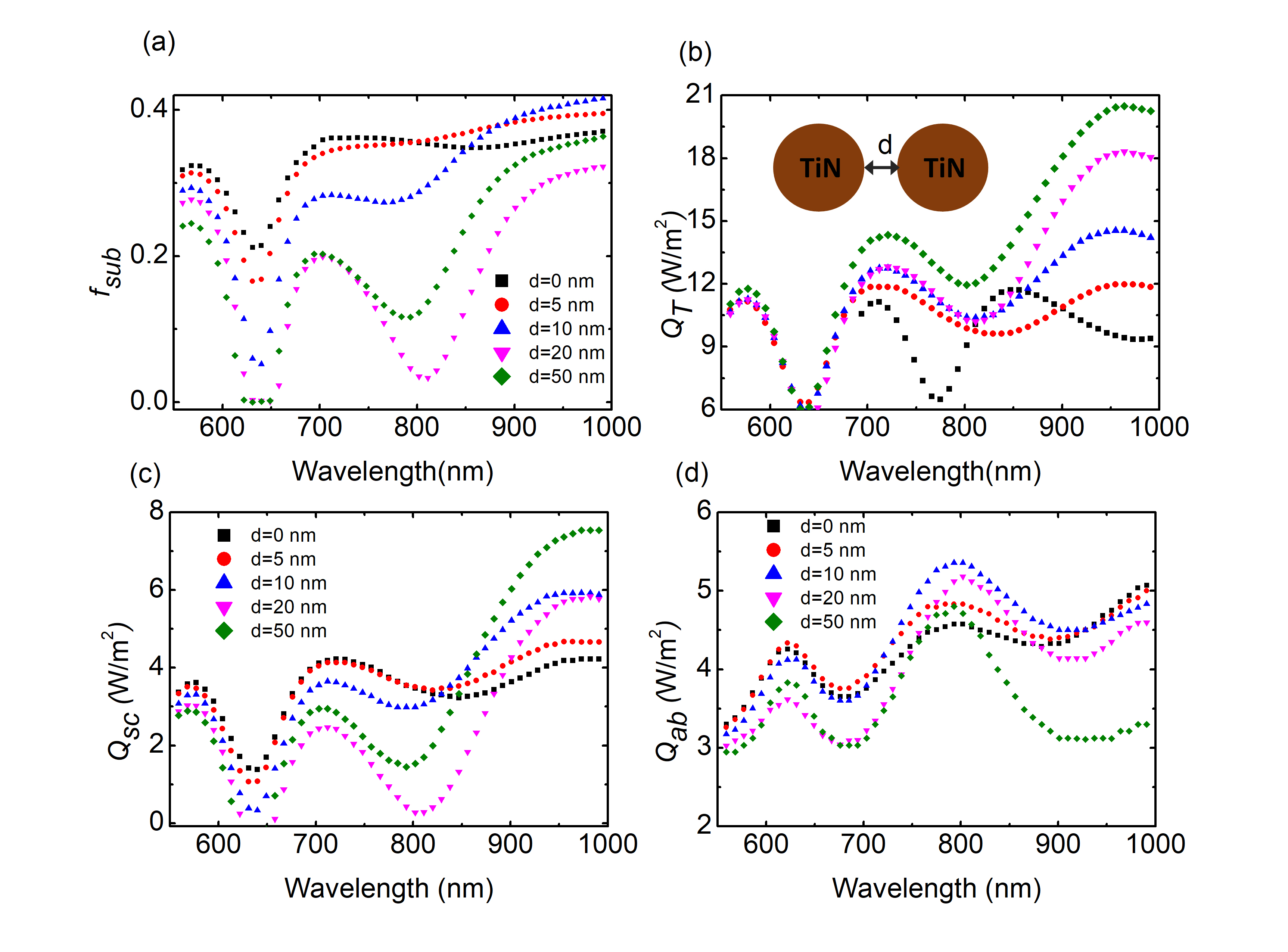}
\caption {(a) $f_{sub}$, (b) $Q_{T}$, (c) $Q_{sc}$, and (d) $Q_{ab}$ as a function of wavelength for TiN dimer spherical NP with distance, d = 0, 20, 70, and 100 nm placed on top of a 30 nm Si$_3$N$_4$ on a Si substrate.}
\label{distance}
\end{figure}

\begin{figure}[h]
\hspace*{-2.cm}
\includegraphics[trim={0cm .5cm 0cm .4cm},clip,width=1.25\textwidth]{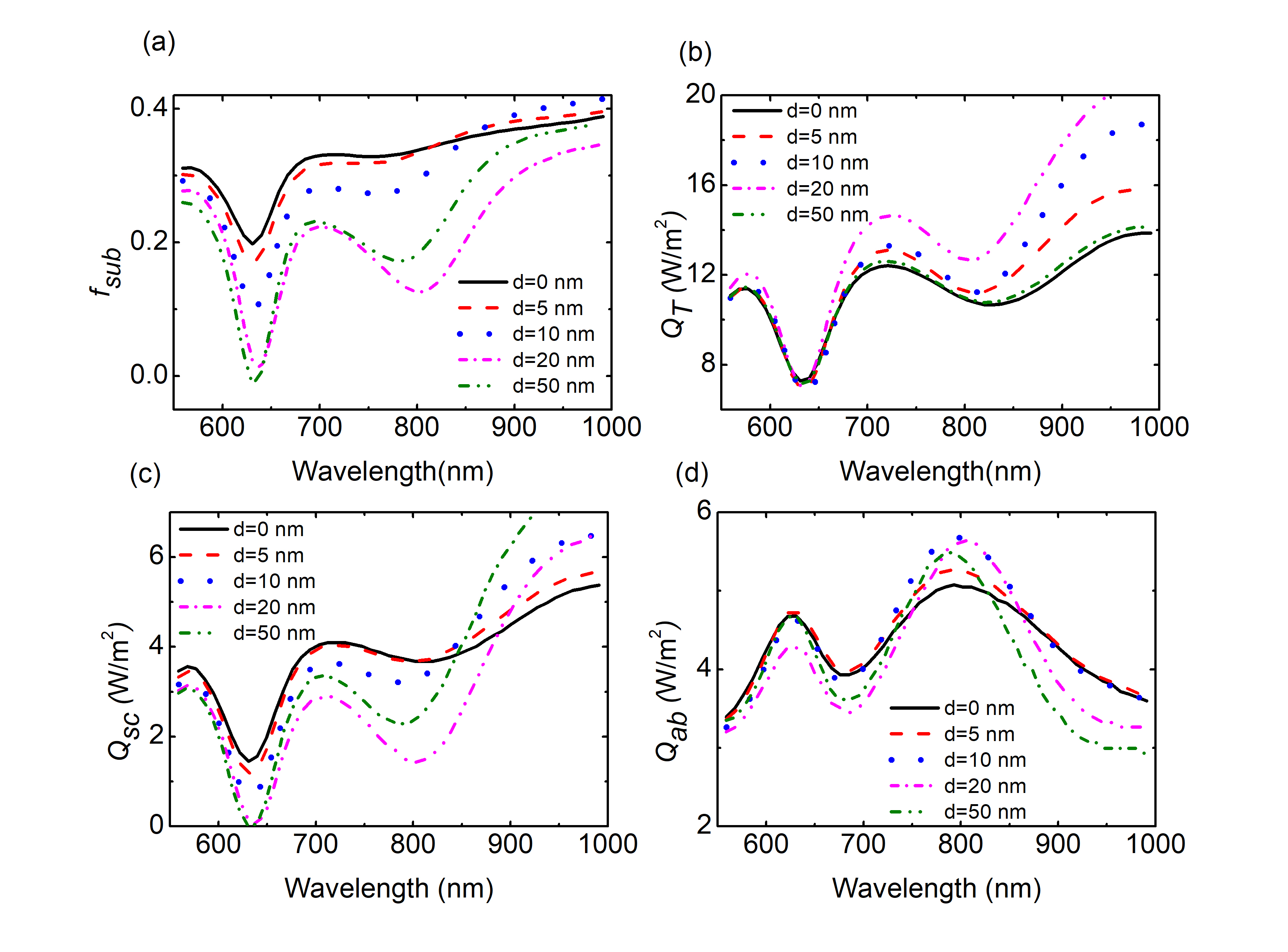}
\caption {(a) $f_{sub}$, (b) $Q_{T}$, (c) $Q_{sc}$, and (d) $Q_{ab}$ as a function of wavelength for TiN dimer spherical NP for 30$^\circ$ the polarization angle of the source with distance, d = 0, 20, 70, and 100 nm placed on top of a 30 nm Si$_3$N$_4$ on a Si substrate.}
\label{distance_angle}
\end{figure}

We considered the impact of changing the distance between the spheres of TiN dimer NPs for the light source polarization angle at 0$^\circ$ and 30$^\circ$. We simulated dimer spherical NPs with various distances between the spheres for $\theta$=0$^\circ$ as can be seen from Fig.\,\ref{distance}. We varied the distance, d from 0 nm to 50 nm to determine the structure for the optimized scattering cross-section. Here, t$_1$ and t$_2$ were considered 30 nm and 250 nm, respectively. The $f_{sub}$ decreased with the increase of the d from d = 0 nm to 20 nm for 550 nm to 850 nm. When d = 50 nm the distance between the dimers increased, and they started to behave like a single sphere, As a result,  $f_{sub}$ increased. When d = 0 nm, the $Q_{T}$ spectrum was the lowest and increased as d increased. The $Q_{sc}$ decreased as d increased from 0 to 20 nm for the 550 nm to 820 nm range. For longer wavelengths than 820 nm, $Q_{sc}$ increased with the increase of d. $Q_{ab}$ was highest when d = 10 nm. For the wavelength 550 to 750 nm, the  $Q_{ab}$ decreased with the increase of d. For the dimers with d = 50 nm, as the distance increased, it started to behave like two independent monomer nanospheres.

We simulated a spherical dimer NP for 30$^\circ$ the polarization angle of the source varying the distance between the spheres as can be seen in Fig.\,\ref{distance_angle}. Here, t$_1$ and t$_2$ were considered 30 nm and 250 nm, respectively. The $f_{sub}$ and $Q_{sc}$ were comparatively higher for the whole spectral range for d = 0 nm and lowest for d=20 nm. When d was smaller than 50 nm the $f_{sub}$ and $Q_{sc}$ increased with the decrease of d, and $Q_{T}$  increased with the increase of d.
The $Q_{T}$ was highest for d = 10 nm and performed better for the wavelength range 740 to 880 nm. Dimers performed better when there was no distance between the spheres. For the polarization angle from 0$^\circ$ to 30$^\circ$, the scattering spectra red-shifted.

\subsection{Dependency on the radius of the dimer spherical NP}

\begin{figure}[h]
\hspace*{-2.cm}
\includegraphics[trim={0cm .5cm 0cm .4cm},clip,width=1.25\textwidth]{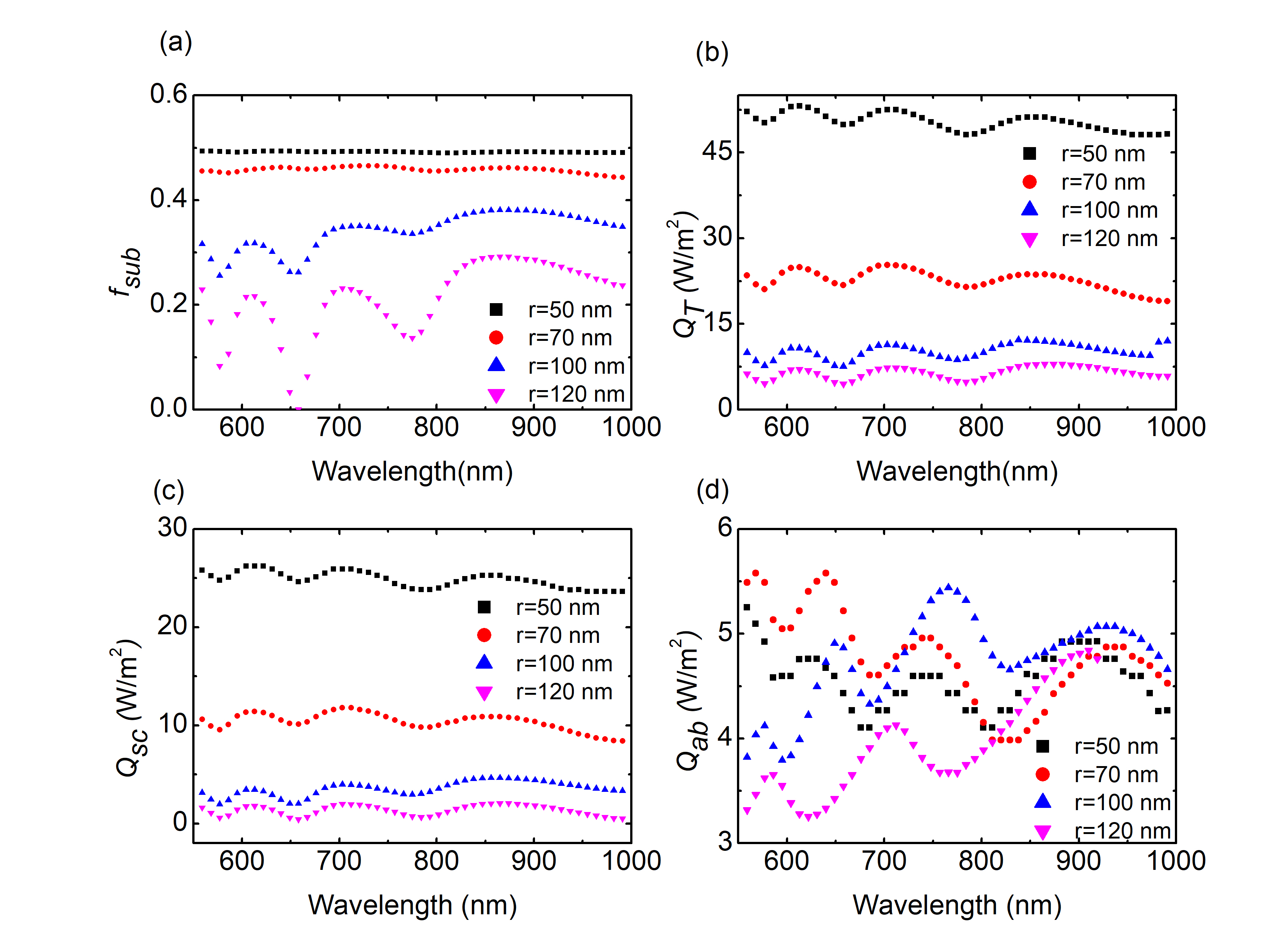}
\caption {(a) $f_{sub}$, (b) $Q_{T}$, (c) $Q_{sc}$, and (d) $Q_{ab}$ as a function of wavelength for dimer spherical TiN NPs with radius, r = 50, 70, 100, and 120 nm placed on top of a 30 nm Si$_3$N$_4$ on a Si substrate.}
\label{radius}
\end{figure}

We simulated spherical dimer NPs with various radii and observed their optical characteristics, as seen from Fig.\,\ref{radius}. To find the best structure for scattering cross-section, we varied the r from 50 nm  to 120 nm. When the radius was 50 nm, $Q_{T}$ was highest and the peak was almost 50 W/m$^2$. 

The most important element to consider when modeling a light-trapping structure's increased path length is $f_{sub}$\,\cite{bohren2008absorption}. As the radius increased, the value of $f_{sub}$ decreased significantly \cite{Akhtary2023}. For the whole spectral range, $f_{sub}$ did not vary appreciably at 50 nm and 70 nm radius. This made it possible to efficiently couple the part of the scattered light with a high in-plane wave vector that is transient in  the air but engendered in silicon. As can be seen in Figs.\,\ref{radius}(b)-(c), $Q_{T}$ and the $Q_{sc}$ decreased as the r increased from 50 nm to 120 nm. $Q_{ab}$ exhibited the highest value for the whole spectral range for r = 100 nm and the peak was at 5.5 W/m$^2$ as can be seen from Figs.\,\ref{radius}(d). Therefore 100 nm radius is optimum for dimers applications.

\section{Conclusions}
Dimer of spherical TiN NPs appeared as an efficient alternative plasmonic material for plasmonic and metamaterial applications.The results of our investigation into the effect of TiN dimer spherical NPs on the enhancement of the thin solar cell's light absorption were promising. After the incorporation of TiN NPs on a silicon substrate, the average absorbed power increased significantly from $\sim$19$\%$ to $\sim$75$\%$ over the whole spectral range. TiN exhibited better absorption enhancement, $g$ and percentage absorbed power to Ag, Au, and Al dimers for r = 15. The average enhancement, G for TiN, Au, and Ag were found to be 0.9972, 0.9953. and 0.9954,  respectively, for r = 15 nm. TiN dimer NP had the highest $Q_{ab}$ value of $\sim$6.2 W/m$^2$ which were greater than Ag, Au, and Al. By changing the size of TiN dimer NPs, the absorption enhancement peak may be tailored to the required solar spectrum. TiN dimer NPs demonstrated to be beneficial when inserted in tandem solar cells because of their cost-effectiveness along with their abundance and ease to manufacture.

\begin{backmatter}
\bmsection{Funding}
A.Z acknowledges the Basic Research Grant (Sonstha/R-60/Ref-4747) provided by the Bangladesh University of Engineering and Technology.

\bmsection{Acknowledgments}
N.A. and A.Z. acknowledge the technical support of the Department of Electrical and Electronic Engineering at Bangladesh University of Engineering and Technology (BUET), Dhaka, Bangladesh, for the completion of the work. 

\bmsection{Disclosures}
The authors declare no conflict of interest.

\bmsection{Data availability}
 Data underlying the results presented in this paper are not publicly available at this time but may be obtained from the authors upon reasonable request.
 
 \bmsection{Supplemental document}
 See Supplement 1 for supporting content.

\end{backmatter}
\bibliography{Dimer_NP}
\end{document}